\documentclass[journal=jpccck,manuscript=article]{achemso}

\usepackage[version=3]{mhchem} 
\usepackage{epsfig}
\usepackage{amsmath}
\usepackage{gensymb}

\let\oldAA\AA
\renewcommand{\AA}{\text{\normalfont\oldAA}}

\author{Freerk Sch\"utt}
\affiliation{Carl-von-Ossietzky Universit{\"a}t Oldenburg, Institute of Physics, 26129 Oldenburg, Germany}
\author{Ana M. Valencia}
\affiliation{Carl-von-Ossietzky Universit{\"a}t Oldenburg, Institute of Physics, 26129 Oldenburg, Germany}
\altaffiliation{Humboldt-Universit\"at zu Berlin, Physics Department and IRIS Adlershof, 12489 Berlin, Germany}
\author{Caterina Cocchi}
\affiliation{Carl-von-Ossietzky Universit{\"a}t Oldenburg, Institute of Physics, 26129 Oldenburg, Germany}
\alsoaffiliation{Carl-von-Ossietzky Universit{\"a}t Oldenburg, Center for Nanoscale Dynamics, 26129 Oldenburg, Germany}
\altaffiliation{Humboldt-Universit\"at zu Berlin, Physics Department and IRIS Adlershof, 12489 Berlin, Germany}
\email{caterina.cocchi@uni-oldenburg.de}

\title{First-Principle Characterization of Structural, Electronic, and Optical Properties of Tin-Halide Monomers}

\begin{document}

\begin{abstract}
The growing interest in tin-halide semiconductors for photovoltaic applications demands an in-depth knowledge of the fundamental properties of its constituents, starting from the smallest monomers entering the initial stages of formation. In this first-principles work based on time-dependent density-functional theory, we investigate the structural, electronic, and optical properties of tin-halide molecules \ce{SnX_n^{2-n}}, with $n=1,2,3,4$ and X = Cl, Br, I, simulating these compounds in vacuo as well as in an implicit solvent. We find that structural properties are very sensitive to the halogen species while the charge distribution is also affected by stoichiometry. The ionicity of the Sn-X bond is confirmed by the Bader charge analysis albeit charge displacement plots point to more complex metal-halide coordination. Particular focus is posed on the neutral molecules \ce{SnX2}, for which electronic and optical properties are discussed in detail. Band gaps and absorption onset decrease with increasing size of the halogen species and despite general common features, each molecule displays peculiar optical signatures. Our results are elaborated in the context of experimental and theoretical literature, including the more widely studied lead-halide analogs, aiming to contribute with microscopic insight to a better understanding of tin-halide perovskites.
\end{abstract}

\newpage

%
%




\section*{Introduction}
Due to their many practical applications and enticingly complex bonding mechanisms, tin halides have been the subject of experimental and theoretical research for decades.
Early on, research was primarily focused on their use in electric discharge lamps~\cite{work1981chemistry,hirayama-straw1984mass}, but has since shifted to applications in the semiconductor industry~\cite{park-etal1986halide,haynes-etal2008laser,hendricks-etal1998metal,geohegan-eden1984column}.
Recently, scientific efforts dedicated to tin halides and their derivatives have reached new heights due to the remarkable performance of their perovskite structures in optoelectronic applications~\cite{heo-etal2021enhancing,abate2023stable,aktas-etal2022challenges}.
Tin halide perovskites have proven to be the most promising substitutes for lead halide perovskites~\cite{nasti-abate2020tin,abate2023stable,aktas-etal2022challenges}: the latter continually astound with ever-increasing performance~\cite{lye-etal2023review,huang-etal2023substitution}, especially regarding the power conversion efficiency of thin film photovoltaics~\cite{green-etal2023solar,duan-etal2023stability}, but remain problematic due to their inherent toxicity~\cite{who2021exposure,who1995inorganic,nasti-abate2020tin}.
In the effort to close the performance gap to the superior Pb-based devices, various tin-halide perovskite structures and their precursors have been studied with a variety of techniques~\cite{jiang-etal2021one,nasti-abate2020tin,digirolamo-etal2021solvents,pascual-etal2022lights,chung-etal2012cssni3,ikram-etal2022recent,dalpian-etal2017changes}.
These complementary experimental and computational studies have led to a dramatic increase in device performances~\cite{green-etal2023solar,ikram-etal2022recent}, based on an improved understanding of their fundamental properties~\cite{aktas-etal2022challenges,abate2023stable,nasti-abate2020tin,abdel-shakour-etal2021high} and their synthesis~\cite{schuett-etal2023electronic,radicchi-etal2022solvent,pascual-etal2020origin,pascual-etal2022lights}, mirroring the early development of lead halide perovskites.

Despite this ever-growing interest in complex tin-halide structures, fundamental research on their molecular units has been stagnating since the 2000s~\cite{hargittai2000molecular,hargittai2009structural,neizer-etal2007vapor,kolonits-etal2004molecular,levy-etal2003structure}, when computational chemistry had a breakthrough.
Therefore, only a few computational studies of these molecules exist~\cite{levy-etal2003structure,kolonits-etal2004molecular,neizer-etal2007vapor,benavides-garcia-balasubramanian1994bond,ramondo-etal1989molecular,ricart-etal1986molecular} and the available ones are focused on their structural and thermodynamic properties~\cite{levy-etal2003structure,kolonits-etal2004molecular,neizer-etal2007vapor}, which could be compared directly to experimental results~\cite{kolonits-etal2004molecular,hargittai2000molecular,hargittai2009structural,gershikov-etal1986combined,lister-sutton1941investigation,demidov-etal1983electron}.
This has created a gap in knowledge abouton the fundamental quantum -mechanical properties of these systems, which is detrimental to the development and understanding of tin -halide perovskite semiconductors.
For example, several open questions remain regarding the characteristics of tin -halide bonds, which have been shown to cause stability problemsissues~\cite{jiang-etal2021one,xiao-etal2019lead,xie-etal2019insight}.
Likewise, the possibility of substituting iodine with lighter halogen species has been only superficially explored in tin-based perovskites~\cite{fu-etal2018organic,bala-kumar2019role,zhou-etal2017low,xu-etal2023synthesis,sahoo-etal2023electrical} although it could represent a viable way to enhance the stability and the performance of these systems.
Understanding the effects of this substitution at the level of tin-halide monomers will provide valuable insight to guide corresponding attempts in more complex compositions, for example, by mixing halide species~\cite{tao-etal2017accurate, tao-etal2019absolute, walsh2015principles}.

In the framework of time-dependent density functional coupled with the polarizable continuum model, we present a comprehensive first-principle investigation of tin-halide molecules with chemical formula \ce{SnX_n^{2-n}}, with $n=1,2,3,4$ and X = Cl, Br, I.
Calculated bond lengths and angles are discussed and compared to experimental results, and the charge distribution across the tin-halide bonds is assessed through atomic partial charges and charge displacement analysis.
Electronic and optical properties are discussed with a focus on the charge-neutral species \ce{SnX2}.
The analysis of the energy levels in vacuo and in solution is supported by the visualization of the molecular orbitals (MOs).
In the discussion of the absorption spectra, emphasis is put on the role of the halogen species in determining the intensity and composition of the optical excitations.
Our results shed light on the characteristics of tin halide monomers providing relevant information for the formation of corresponding semiconductor films and essential insight into their fundamental properties.

\section*{Methodology}
\subsection*{Model Systems}
\begin{figure}
	\begin{center}
		\includegraphics[width=\textwidth]{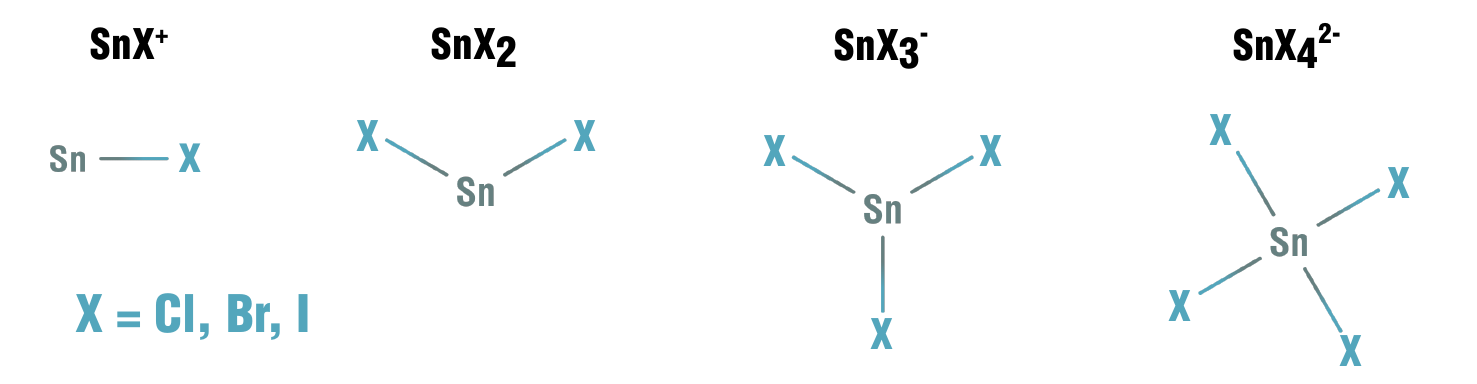}
	\end{center}
	\caption{Structural formula of the four \ce{SnX_n^{2-n}} molecules considered in this work.}
	\label{fig:StructureOverview}
\end{figure}

In this study, we consider tin-halide molecules with chemical formula \ce{SnX_n^{2-n}} where $n=1...4$ and X = Cl, Br, I (see Figure~\ref{fig:StructureOverview}).
Out of these structures, the tin dihalides (\ce{SnX2}) and tin tetrahalides (\ce{SnX4^2-}) species are the most common ones and have been shown to exist in the vapor phase both in the monomeric and dimeric form~\cite{hargittai2000molecular,levy-etal2003structure,kolonits-etal2004molecular,neizer-etal2007vapor}.
Here, we specifically focus on the monomers as they have been shown to prevail over dimers~\cite{hargittai2000molecular}.
Moreover, we pose particular emphasis on the charge neutral molecules \ce{SnX2}, as they represent the most common coordination of tin halides~\cite{hargittai2000molecular} and are present in tin-halide perovskite structures.

\subsection*{Computational Framework}
The results presented in this study are calculated from first principles using density functional theory (DFT)~\cite{hohenberg-kohn1964inhomogeneous,kohn-sham1965self} as implemented in the software suite Gaussian 16~\cite{frisch-etal2016gaussian}.
To mimic the electrostatic interactions with a surrounding solvent, the polarizable continuum model (PCM)~\cite{miertus-etal1981electrostatic,amovilli-etal1998recent} is adopted.
In all these calculations, an implicit dimethyl sulfoxide (DMSO) solution is assumed having checked that the specific solvent polarity does not cause any qualitative changes in the electronic and optical properties of the tin-halide molecules.

The generalized gradient approximation in the Perdew-Burke-Ernzerhof (PBE) parameterization~\cite{perdew-etal1996generalized} is chosen for the exchange-correlation functional for the relaxation of the \ce{SnX_n^{2-n}} monomers, as it offers a good balance between accuracy and computational efficiency and has been successfully adopted in studies of equivalent Pb-halide molecules and metal-halide systems in general~\cite{radicchi-etal2019understanding,schier-etal2021formation,valencia-etal2021optical,procida-etal2021first,kaiser-etal2021iodide,schuett-etal2023electronic}.
The semi-empirical Grimme-D3 dispersion scheme is employed to account for van der Waals interactions~\cite{grimme-etal2010consistent}.
The relaxed structures are verified to be actual minima using vibrational frequency calculations.
All subsequent single-point geometry, population analysis, and time-dependent DFT (TDDFT) calculations~\cite{casida1996time,runge-gross1984density} are performed on the relaxed molecules with the range-separated hybrid functional CAM-B3LYP~\cite{yanai-etal2004new}.
In these runs, the SDD basis set and pseudopotential are used for the Sn and I atoms, while Cl and Br are modeled with the cc-pVQZ basis set.
Open-shell molecules are treated in a spin-unrestricted framework.
The Bader charge analysis is performed according to the ``atoms in molecules'' (AIM) method~\cite{tang-etal2009grid} implemented in the Multiwfn software~\cite{lu-chen2012multiwfn}.

The analysis of the \ce{SnX2} molecules is supplemented by the calculation of the charge displacement (CD) upon the formation of the Sn-X bond between the \ce{SnX} and \ce{X} fragments.
The CD analysis~\cite{belpassi-etal2008chemical} has been successfully used to describe the nature of coordinating bonds and especially the influence of interatomic charge transfer in bond formation~\cite{belpassi-etal2009experimental, belpassi-etal2010charge, borghesi-etal2019nature, cappelletti-etal2011nature, nunzi-etal2019insight}.
Moreover, it has been shown to produce meaningful results for similar halogen bonds in comparison to experiments~\cite{cappelletti-etal2011nature,borghesi-etal2019nature, nunzi-etal2019insight}.
The CD analysis is based on the difference $\Delta\rho$ between the electronic density of the total molecule and the density of the two non-interacting fragments~\cite{belpassi-etal2008chemical, nunzi-etal2019insight, borghesi-etal2019nature}.
Subsequently, the CD function $\Delta Q$ is calculated as a partial integral of $\Delta\rho$ along the axis of the Sn-X bond:
\begin{equation}
	\Delta Q(z) = \int_{-\infty}^\infty \text{d}x \int_{-\infty}^\infty \text{d}y \int_{-\infty}^z \Delta\rho(x,y,z') \text{d}z'.
	\label{eq:CTFunction}
\end{equation}
In the adopted framework, the interatomic Sn-X axis is aligned along the $z$-axis.
The calculated charge displacement $\Delta Q$ provides quantitative information about the charge distribution in the bond by measuring the amount of electronic charge that is displaced from the right side of a plane perpendicular to the $z$-axis to its left side at the given point.
Positive values of $\Delta Q$ indicate charge displacement of equivalent magnitude from right to left (towards $-z$) and negative $\Delta Q$ identify the corresponding charge displacement from left to right (towards $z$).
Since the difference between any two values of $\Delta Q$ gives the net flux of electronic density into the area between the two points, the CD function also clearly identifies regions of increasing and decreasing charge; with a positive (negative) slope defining an area of charge accumulation (depletion).
Since this method necessitates an accurate description of the electronic density of the system, including core electrons, for this analysis performed in vacuo we used the all-electron code FHI-aims~\cite{blum-etal2009ab,blum-etal2022fhi} with the PBE functional~\cite{perdew-etal1996generalized} and tight basis sets~\cite{havu-etal2009efficient}.

\section*{Results and Discussion}

\subsection*{Structural Properties}

In the analysis of the structural properties, we start from the \ce{SnX2} molecules which were constructed considering two different starting configurations: one geometry with a 90\textdegree{} Sn-X-Sn angle, known as the ``axial configuration'', with $\text{C}_\text{2v}$ symmetry, and one with a 180\textdegree{} Sn-X-Sn angle, called ``linear configuration'', with $\text{C}_{\infty\text{h}}$ symmetry.
Between them, only the axial arrangement of the molecules converged to a real minimum in the potential energy surface, while the linear geometries converged to higher energy saddle points.
This finding is in agreement with previous experimental and computational studies of group 14 metal dihalides, which conclusively showed that the $\text{C}_\text{2v}$ symmetry is the most favorable arrangement for these molecules~\cite{demidov-etal1983electron,lister-sutton1941investigation,hargittai2000molecular,neizer-etal2007vapor,kolonits-etal2004molecular,levy-etal2003structure}.
The higher stability of \ce{SnX2} molecules in their axial configuration can be further rationalized considering the additional lone electron pair in the valence shell of Sn~\cite{hargittai-chamberland1986vsepr}.
Based on these findings, these relaxed geometries of the \ce{SnX2} compounds are adopted in the subsequent steps of this study.

\begin{figure}[h!]
	\begin{center}
		\includegraphics[width=0.85\textwidth]{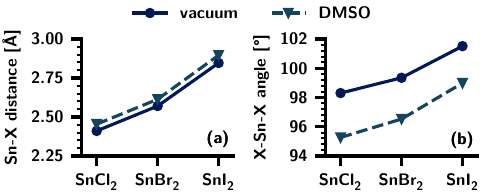}
	\end{center}
	\caption{Structural properties of the \ce{SnX2} molecules in vacuum (circles) and in an implicit DMSO solution (triangles), including (a) the average Sn-X distance, and (b) the X-Sn-X angle, with X = Cl, Br, I.}
	\label{fig:SnX2StructuralProperties}
\end{figure}

Figure~\ref{fig:SnX2StructuralProperties} reports the distances between the Sn and the halogen atoms X (Sn-X distance) and the X-Sn-X angle for the three halide species, both with the implicit solvent DMSO and in vacuo.
The exact metrics of the calculated structures are shown in Table~\ref{tab:SnX2structureMetrics}.
The larger halide species (Br and I) give rise to longer Sn-X distances and a wider X-Sn-X angle, showing a monotonic increase of approximately 0.4~\AA{} and 4\textdegree{} with Br and I, respectively.
This clearly identifiable trend in terms of both the bond distances and angle is a common feature of almost all metal halides~\cite{hargittai2000molecular,hargittai2009structural,ricart-etal1986molecular}.
The electrostatic interaction with the implicit solvent causes a further increase inof the Sn-X separation and a decrease in the X-Sn-X angle.
Furthermore, while the increment in the Sn-X distance is only of the order of 0.05~\AA{}, the narrowing of the angle due to the solvent is more significant, exhibiting a decrease compared to the molecule in vacuo of 3.0\textdegree{}, 2.8\textdegree{}, and 2.5\textdegree{} with X = Cl, Br, and I, respectively.
Overall, the smallest bond distances and angles are obtained for \ce{SnCl2} with 2.41~\AA{} (2.45~\AA{}) and 98.3\textdegree{} (95.3\textdegree{}) in vacuum (in DMSO), while the largest metrics pertain to \ce{SnI2} with 2.85~\AA{} (2.9~\AA{}) and 101.5\textdegree{} (99\textdegree{}), as shown in Figure~\ref{fig:SnX2StructuralProperties} and summarized in Table~\ref{tab:SnX2structureMetrics}.

\begin{table}[h]
	\small
	\caption{Sn-X distances and X-Sn-X angle of the \ce{SnX2} molecules in vacuo and in an implicit DMSO solution.}
	\label{tab:SnX2structureMetrics}
	\begin{tabular*}{\textwidth}{@{\extracolsep{\fill}}lcccc}
		\hline
		\multicolumn{1}{c}{} & \multicolumn{2}{c}{in vacuo} & \multicolumn{2}{c}{in DMSO}  \\
		\hline
		&Sn-X [\AA{}]  &X-Sn-X [\textdegree]  &Sn-X [\AA{}] &X-Sn-X[\textdegree] \\\hline
		\ce{SnCl2}           & 2.41    & 98.3     & 2.45        & 95.3                   \\
		\ce{SnBr2}           & 2.57                       & 99.3                     & 2.61        & 96.5                   \\
		\ce{SnI2}            & 2.85                       & 101.5                    & 2.90        & 99.0                   \\
		\hline
	\end{tabular*}
\end{table}

It is instructive to compare our computational results for bond length and distances of the considered \ce{SnX2} molecules with the available computational~\cite{benavides-garcia-balasubramanian1994bond,ricart-etal1986molecular,levy-etal2003structure,kolonits-etal2004molecular,neizer-etal2007vapor,szabados-hargittai2003molecular,elicker-etal2000molecular} and experimental references~\cite{demidov-etal1983electron,lister-sutton1941investigation,kolonits-etal2004molecular,gershikov-etal1986combined,nasarenko-etal1985second,ermakov-etal1991application}.
The calculated Sn-Cl distances of 2.41~\AA{} in vacuo and of 2.45~\AA{} in DMSO differ by a few picometers from most theoretical~\cite{benavides-garcia-balasubramanian1994bond,szabados-hargittai2003molecular,elicker-etal2000molecular} and experimental~\cite{nasarenko-etal1985second,ermakov-etal1991application,gershikov-etal1986combined} values ranging between 2.33~\AA{} and 2.39~\AA{}.
Remarkably, an old experimental result of 2.42~\AA{} determined with electron diffraction~\cite{lister-sutton1941investigation} is very close to our results.
Likewise, in \ce{SnBr2}, reference values calculated with various computational methods~\cite{benavides-garcia-balasubramanian1994bond,szabados-hargittai2003molecular,elicker-etal2000molecular} and experimental values determined using electron diffraction~\cite{demidov-etal1983electron,lister-sutton1941investigation,nasarenko-etal1985second,ermakov-etal1991application} range between 2.50~\AA{} and 2.55~\AA{} and are thus a few picometers smaller than our calculated Sn-Br distances of 2.57~\AA{} in vacuo and 2.61~\AA{} in DMSO.
For \ce{SnI2}, the variations of our results compared to the literature are slightly larger: while we obtained an Sn-I separation of 2.85~\AA{} in vacuo and 2.90~\AA{} in DMSO, available theoretical references~\cite{neizer-etal2007vapor,benavides-garcia-balasubramanian1994bond,szabados-hargittai2003molecular,elicker-etal2000molecular} and experimental data~\cite{demidov-etal1983electron,lister-sutton1941investigation,nasarenko-etal1985second,ermakov-etal1991application} report values between 2.71~\AA{} and 2.78~\AA{}.

Discrepancies between calculated and experimental metal-halogen distances are a known issue in computational chemistry~\cite{hargittai2000molecular,szabados-hargittai2003molecular}
which can be attributed to the unusual nature of the corresponding bonds and to the size of the atoms involved.
Some studies showed that the most common basis sets for DFT methods are often unable to fully capture the electronic configuration in bonding processes of large atoms, such as Sn and I~\cite{hargittai-varga2007molecular,hargittai2000molecular}.
Relativistic effects are known to play a role, too~\cite{hargittai2000molecular,schwerdtfeger-etal1992relativistic}.
Finally, second-order Moller-Plesset (MP2) or coupled cluster single-double(-triplet) methods~\cite{hargittai2000molecular,hargittai2009structural,neizer-etal2007vapor,levy-etal2003structure,kolonits-etal2004molecular} are proven to be superior to DFT in the calculation of metal halide structures.
Nonetheless, the trend provided by our study can be considered qualitatively and quantitatively reliable.

The X-Sn-X angles calculated in this work are in good agreement with published experimental values determined with electron diffraction and vibrational corrections: the results of 98.3 and 99.3\textdegree obtained for \ce{SnCl2} and \ce{ScBr3} in vacuo compare very well with the available references of 98.1\textdegree \cite{levy-etal2003structure,nasarenko-etal1985second} and 99.7\textdegree \cite{kolonits-etal2004molecular,demidov-etal1983electron}, respectively.
On the other hand, for \ce{SnI2}, we obtained 101.5\textdegree (see Table~\ref{tab:SnX2structureMetrics}) while the literature reports 103.5\textdegree \cite{neizer-etal2007vapor,nasarenko-etal1985second}.
The values of tin-halogen bond angles reported in earlier computational references~\cite{neizer-etal2007vapor,benavides-garcia-balasubramanian1994bond,szabados-hargittai2003molecular,elicker-etal2000molecular} are all in excellent agreement with those reported in this work.

\begin{figure}[h!]
	\begin{center}
		\includegraphics[width=0.67\textwidth]{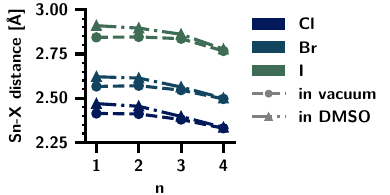}
	\end{center}
	\caption{Average Sn-X distance of the considered \ce{SnX_n^{2-n}} molecules with X = Cl, Br, I and $n=1...4$ computed in vacuum (circles) and in DMSO (triangles).}
	\label{fig:SnXnTinHalideDistances}
\end{figure}

The overall structural trends obtained for \ce{SnX2} hold for the other \ce{SnX_n^{2-n}} molecules, although the charged monomers possess their distinct structural characteristics (see Figure~\ref{fig:StructureOverview}).
The optimized cations \ce{SnX+} assume a linear configuration with $\text{C}_{\infty\text{h}}$ symmetry,  while for the single and double anions, \ce{SnX3-} and \ce{SnX4^2-} respectively, two initial arrangements are considered:
\ce{SnX3-} are relaxed from a trigonal pyramidal ($\text{C}_{\text{3v}}$) and a trigonal planar geometry ($\text{C}_{\text{3h}}$), while \ce{SnX4^2-} from a tetrahedral ($\text{T}_\text{d}$) and a square planar geometry ($\text{D}_{\text{4h}}$).
For both molecules, the non-planar arrangements are more stable and thus are assumed as the equilibrium geometries.
As shown in Figure~\ref{fig:SnXnTinHalideDistances}, the increase of $n$ causes a monotonic decrease of 0.05~\AA{} in the average Sn-X distances, ranging from approximately 2.40~\AA{} ($n=1$) to 2.35~\AA{} ($n=4$) in the presence of Cl; from about 2.55~\AA{} ($n=1$) to 2.50~\AA{} ($n=4$) with Br; from approximately 2.85~\AA{} ($n=1$) to less than 2.80~\AA{} ($n=4$) with I.
Electrostatic solvent interactions cause a slight enlargement of the Sn-X distance on the order of 0.05~\AA{}.
However, this effect is negligible with $n \geq 3$ where the values computed in vacuo and in solution essentially overlap (Figure~\ref{fig:SnXnTinHalideDistances}).
The Sn-X distances reported in this work are consistent with available theoretical and experimental references~\cite{demidov-etal1983electron,lister-sutton1941investigation,nasarenko-etal1985second,ermakov-etal1991application,neizer-etal2007vapor,benavides-garcia-balasubramanian1994bond,szabados-hargittai2003molecular,elicker-etal2000molecular,gershikov-etal1986combined} and this agreement includes the reduction at increasing $n$~\cite{elicker-etal2000molecular,nakamoto1986infrared,fujii-kimura1970molecular,escalante-etal1999structure,escalante-etal1999structure,fonda-etal2020mechanism} discussed above.

Before closing this section, it is worth comparing the structural parameters obtained for the herein-considered tin dihalide monomers to results obtained for their \ce{PbX2} counterparts in theoretical and experimental studies.
In agreement with the observed trends for group-14 metal halides of decreasing bond lengths with larger metal and halogen atoms~\cite{hargittai2000molecular}, results from electron diffraction measurements~\cite{hargittai-etal1977two,bazhanov1991structure,gershikov-etal1986combined} and calculated equilibrium Pb-X bond lengths~\cite{benavides-garcia-balasubramanian1994bond,escalante-etal1999structure,howard1994ab,schier-etal2021formation} are consistently smaller than the Sn-X distances reported here.
The same trend is found also for the bond angles which decrease with the increasing weight of the group-14 atoms~\cite{hargittai2000molecular}, except for the lead dihalides, which have significantly larger bond angles than \ce{SnX2}.
This behavior matches our calculated X-Sn-X angles and available reference values for the \ce{PbX2} molecules~\cite{benavides-garcia-balasubramanian1994bond,escalante-etal1999structure,howard1994ab,schier-etal2021formation}.
Irrespective of these quantitative differences, the \ce{SnX2} and \ce{PbX2} molecules are very similar in their bend geometries.
Both tin halides and lead halides are most commonly found in this two-coordination, though \ce{PbX2} is especially stable in this configuration due to the inert pair effect~\cite{hargittai2000molecular,drago1958thermodynamic}.

\subsection*{Charge Analysis}
\begin{figure}[htpb]
	\begin{center}
		\includegraphics[width=0.85\textwidth]{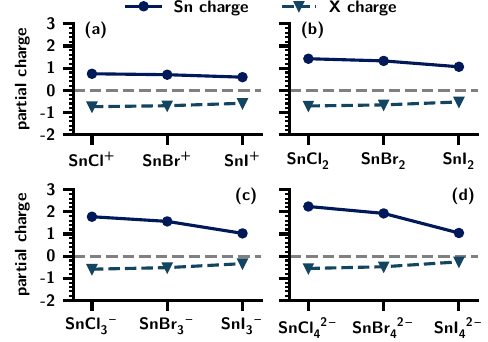}
	\end{center}
	\caption{Bader partial charges on Sn (circles) and X = Cl, Br, I (triangles) in \ce{SnX_n^{2-n}} monomers with a) $n=1$, b) $n=2$, c) $n=3$, and d) $n=4$.
	}
	\label{fig:baderCharges}
\end{figure}

To gain a better understanding of the nature of the Sn-halide bonds in the \ce{SnX_n^{2-n}} molecules, we analyze the Bader partial charges using the AIM framework~\cite{tang-etal2009grid} implemented in the Multiwfn software~\cite{lu-chen2012multiwfn}.
Regardless of the number of halide atoms $n$, the polarization of the Sn-X bonds in the molecules decreases monotonically with the halogen radius, showing a positive charge on Sn and a negative charge on X (see Figure~\ref{fig:baderCharges}).
This clear trend, which was also reported for the equivalent lead-halide molecules~\cite{ramondo-etal1989molecular}, is likely caused by the decreasing electronegativity of the halogen species ranging from 3.5 for Cl to 3.45 for Br down to 3.2 for I~\cite{tantardini-oganov2021thermochemical}, directly opposing the trend reported for \ce{PbX2}~\cite{schier-etal2021formation}.

At increasing $n$, the formation of chemical bonds with a higher number of halogen atoms enhances the positive partial charge of Sn by approximately 0.5 per halogen atom, due to the additional electronic charge transferred away from the metal center.
The resulting charge delocalization across the Sn-halide bonds leads to a reduction of the negative charge on the halogen atoms of about 0.4 from \ce{SnX+} to \ce{SnX4^2-} with similar trends for the charged molecule (Figure~\ref{fig:baderCharges}b) and the cation (Figure~\ref{fig:baderCharges}a) as well as for the mono- and the dianion (compare Figure~\ref{fig:baderCharges}c and d). This behavior suggests an increasingly covalent character of the Sn-X bonds in the larger molecules.
Although the exact values of the atomic partial charges calculated with different methods cannot be compared quantitatively, existing results for lead- and tin-dihalides suggest that the Pb-X bond is generally more polarized than the equivalent Sn-X bond~\cite{borghesi-etal2019nature,schier-etal2021formation,ramondo-etal1989molecular}.

\begin{figure}[htpb]
	\begin{center}
		\includegraphics[width=0.9\textwidth]{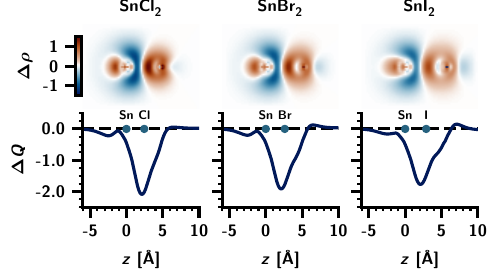}
	\end{center}
	\caption{Charge displacement plots relative to the Sn-X bond for \ce{SnCl2} (left), \ce{SnBr2} (middle), and \ce{SnI2} (right). The top panels display the electronic density difference, $\Delta\rho$, while at the bottom, the corresponding charge displacement function $\Delta Q$ is visualized. 
	}
	\label{fig:SnX2ChargeDisplacement}
\end{figure}

For a deeper understanding of the charge distribution around the Sn-X bonds, we inspect the CD plots reported in Figure~\ref{fig:SnX2ChargeDisplacement}.
The electronic density difference (top) and the CD function (bottom) calculated for the three \ce{SnX2} molecules (X = Cl, Br, I) is shown as a function of the coordinate $z$ along which one of the Sn-X bonds is aligned.
The CD function $\Delta Q(z)$ gives at any point the amount of charge transferred at that point across a plane perpendicular to the $z$-axis from right to left (towards negative values of $z$).
In the density difference plots, green areas indicate charge depletion upon bond formation, corresponding to a negative slope of $\Delta Q(z)$, while red areas indicate charge accumulation together with a positive slope of $\Delta Q(z)$.
Since the electronic density has circular symmetry around the bond axis, the two-dimensional visualization of the electronic density difference $\Delta\rho$ on one plane contains all relevant information (see Figure~S4 for a three-dimensional plot).
Both the CD function and the electronic density difference show a large transfer of electronic charge, supporting the conclusions based on the partial charge analysis.
In fact, except for a small region far behind the halogen atom, the CD function is negative along the entire bond axis, indicating a monotonic charge transfer from left to right, \textit{i.e.}, from the Sn atom toward X.
Together with the splitting of the bond into a region of strong charge depletion on the side of Sn and a region of charge accumulation on the side of the halogen, this finding further supports the dominant ionic nature of the Sn-X bond.

The charge density difference plots show that upon bond formation the charge is moved away from the vicinity of the Sn atom to the region around the halogen atom.
Yet, the charge gradient visualized in the top panel of Figure~\ref{fig:SnX2ChargeDisplacement} suggests that the Sn-X bond is characterized by more complex coordination, although ionicity prevails in the first order.
The decreasing polarization of the Sn-X bond discussed in the partial charge analysis is also evident in the CD function.
With the larger and less electronegative halogen species in the tin-halide compounds, the net amount of charge transferred along the bond axis decreases significantly, from approximately 2.1 in \ce{SnCl2} to about 1.75 in \ce{SnI2}.
The further CD increase in two small regions of opposing charge displacement with Br and I, which is marked by the flattening of the CD curve behind Sn and by increasing positive CD values far behind the halogen atom, additionally supports the previous assumption of the increasing covalent nature of the Sn-X bond with the larger and less electronegative halogen atoms.
Overall, these results confirm that tin-halogen bonding is a complex interplay of electrostatic effects and charge transfer~\cite{torii2019correlation,cavallo-etal2016halogen,huber-etal2013directionality}, which was also previously observed in a computational study of \ce{SnI2}~\cite{borghesi-etal2019nature} and for metal halides in general~\cite{hargittai2000molecular,hargittai2009structural}.

\subsection*{Electronic Properties}
\begin{figure}[htpb]
	\begin{center}
		\includegraphics[width=0.85\textwidth]{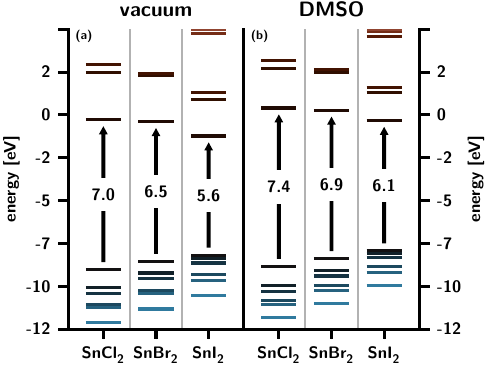}
	\end{center}
	\caption{Energy levels of the \ce{SnX2} molecules (X = Cl, Br, I) in vacuo (a) and in an implicit DMSO solution (b). The HOMO-LUMO gap is highlighted by arrows and the corresponding value is reported in eV.}
	\label{fig:SnX2EnergyLevels}
\end{figure}

We analyze the electronic structure of the charge-neutral molecules \ce{SnX2}, which are the most relevant building blocks of tin-halide perovskites thin films~\cite{hargittai2000molecular}.
The energy eigenvalue spectra computed from these molecules in vacuo and in DMSO are shown in Figure~\ref{fig:SnX2EnergyLevels}.
From the inspection of this graph, we notice that at increasing size of the halogen atoms, (un)occupied orbital energies shift (down)upward, resulting in a notable narrowing of the gap between the highest occupied molecular orbital (HOMO) and the lowest unoccupied one (LUMO).
The same trend was reported for analogous \ce{PbX2} molecules~\cite{ramondo-etal1989molecular} as well as for tin- and lead-halide solution complexes~\cite{schuett-etal2023electronic,procida-etal2021first} and is in agreement with experimental findings on metal-halide perovskites~\cite{tao-etal2017accurate,tao-etal2019absolute,walsh2015principles,xu-etal2023challenges}.

The energy levels below the HOMO and above the LUMO are similarly affected by the halogen atoms in the monomer with some notable differences.
In the occupied region for the molecules in vacuo, the HOMO-1 and HOMO-2 of \ce{SnCl2} and \ce{SnBr2} are almost equally separated from each other by several hundreds of meV lower than the HOMO; in \ce{SnI2}, instead, these two states are much closer to each other and to the HOMO.
The HOMO-3 and HOMO-4 on the other hand, are energetically very close in \ce{SnCl2} and \ce{SnBr2}, while they are separated by a few hundreds of meV in \ce{SnI2}.
In the unoccupied region, the LUMO+1 of \ce{SnCl2} and \ce{SnBr2} is about 2~eV higher in energy than the LUMO; however, in and \ce{SnBr2}, the LUMO+2 is almost overlapping with the LUMO+1 in Figure~\ref{fig:SnX2EnergyLevels}a).
In \ce{SnI2}, the mutual separation between the LUMO+1 and the LUMO+2 is almost identical to the one in \ce{SnCl2} while the LUMO-LUMO+1 distance is a few hundreds of meV lower.
Also, in this molecule, the LUMO+3 and LUMO+4 are visible in Figure~\ref{fig:SnX2EnergyLevels}a whereas their counterparts in the spectrum of \ce{SnCl2} and \ce{SnBr2} are off-scale.
The electronic structure of the \ce{SnX2} molecules is qualitatively similar to that of their \ce{PbX2}~\cite{ramondo-etal1989molecular} counterparts.

The trends for the electronic energy levels discussed for the \ce{SnX2} molecules in vacuo are generally valid also in solution.
However, a careful inspection of Figure~\ref{fig:SnX2EnergyLevels} reveals different responses of the various MOs to the polarizable medium.
Notably, these trends are independent of the halogen species.
The occupied levels displayed in Figure~\ref{fig:SnX2EnergyLevels} undergo a slight downshift of about 100~meV in solution compared to in vacuo.
In contrast, the energies of the unoccupied states sizably increase in DMSO by a few hundred meV.

\begin{figure}[htpb]
	\begin{center}
		\includegraphics[width=\textwidth]{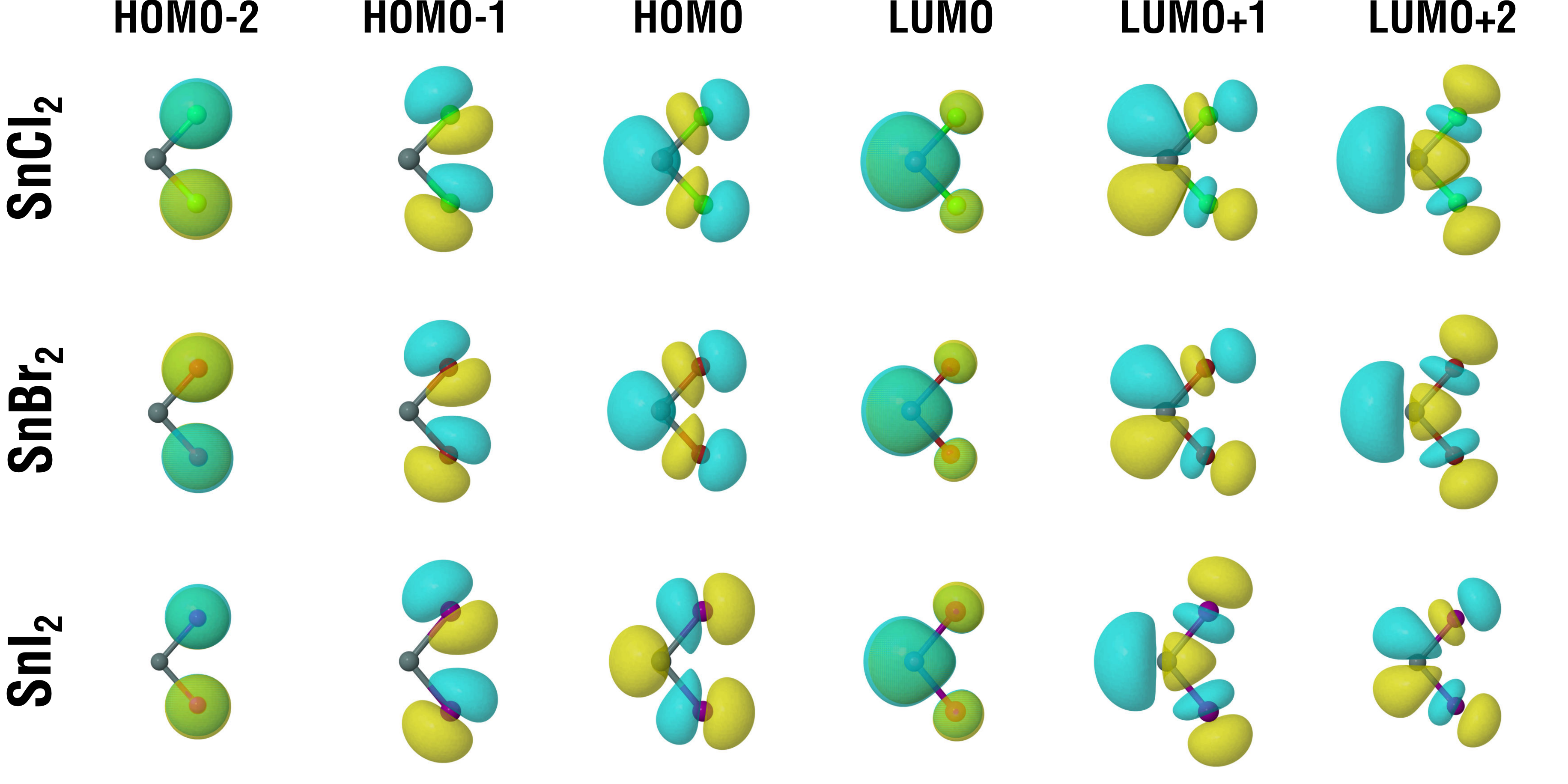}
	\end{center}
	\caption{Spatial distribution of the three highest occupied and lowest unoccupied MOs of \ce{SnX2} with X = Cl, Br, I computed from DFT+PCM. Corresponding results obtained in vacuo are visually identical.
	}
	\label{fig:SnX2MOs}
\end{figure}

To better understand this behavior, we inspect the spatial distribution of the MOs shown in Figure~\ref{fig:SnX2MOs}.
The occupied states are visually identical in the three molecules regardless of the halogen species: the HOMO is formed by a combination of an $s$-state on the Sn atom and a $p$-state on the halogen atoms, while the HOMO-1 and the HOMO-2 are both completely defined by the $p$-orbitals of the halogens with no Sn contributions.
Similar characteristics were found in equivalent lead dihalides~\cite{procida-etal2021first}.
On the other hand, the unoccupied MOs are given by hybridized Sn and halogen $p$-states (see Figure~\ref{fig:SnX2MOs}).
While the character of these MOs is the same in \ce{SnCl2} and \ce{SnBr2}, in \ce{SnI2}, the ordering of the LUMO+1 and the LUMO+2 is swapped compared to the other molecules.
This behavior is mirrored in the lead-based \ce{PbI2} counterparts~\cite{procida-etal2021first} and, thus, cannot be considered a peculiarity of tin-based halide compounds.

\subsection*{Optical Properties}
\begin{figure*}[htpb]
	\begin{center}
		\includegraphics[width=\textwidth]{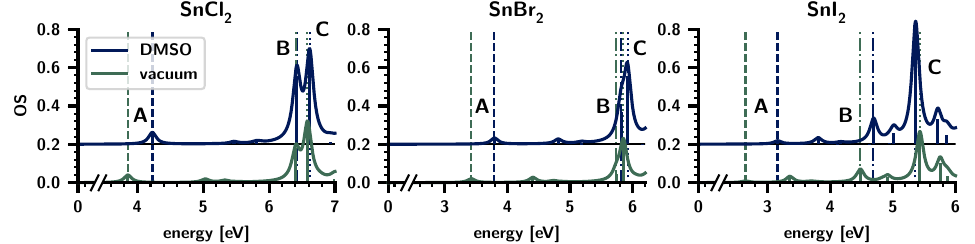}
	\end{center}
	\caption{Optical absorption spectra of \ce{SnCl2} (left), \ce{SnBr2} (middle), and \ce{SnI2} (right), computed in vacuo (green) and in an implicit DMSO solution (blue, vertically offset by 0.2). A Lorentzian broadening of 70 meV is applied to all spectra for visualization purposes. The two excitations discussed in the text are labeled with capital letters and marked by vertical lines. Peak A is marked by a dashed, peak B by a dash-dotted, and peak C by a dotted line.}
	\label{fig:SnX2AbsorptionSpectra}
\end{figure*}

We conclude our study with the analysis of the optical properties of the \ce{SnX2} monomers examining their absorption spectra reported in Figure~\ref{fig:SnX2AbsorptionSpectra}.
In these three systems, the lowest-energy excitation (labeled A), corresponding to the transition between the HOMO and the LUMO, has a low oscillator strength (OS, see Tables S15-S20).
This behavior is readily understood considering the character of the orbitals involved (see Figure~\ref{fig:SnX2MOs}).
The energy of A decreases at increasing size of the halogen atom, consistent with the behavior of the HOMO-LUMO gap (compare Figure~\ref{fig:SnX2EnergyLevels}): in \ce{SnCl2}, its value is close to 4~eV while in \ce{SnI2} it is about 3~eV (see Figure~\ref{fig:SnX2AbsorptionSpectra}).

Above peak A, the spectra of \ce{SnX2} feature two excitations with very low OS, stemming from transitions targeting the LUMO from the HOMO-2 and the HOMO-4.
Since both these occupied states lack the $p$-orbital component on the Sn atom, they do not give rise to a strong transition dipole moment when overlapped with the LUMO.
At even higher energies, the most intense peaks appear. 
In the spectrum of \ce{SnCl2}, they give rise to two sharp peaks, labeled B and C, found at approximately 6.4~eV and 6.6~eV, respectively: the former is an excitation from the HOMO to LUMO+1 while the latter comes from the transition between the HOMO and the LUMO+2.
In the result obtained for \ce{SnBr2}, these excitations, while retaining the same composition, are energetically closer, being separated by only 100 meV.
In the spectrum of tin bromide, peak B is weaker than peak C, due to the slightly reduced overlap between the HOMO and the LUMO+1 (compare Figure~\ref{fig:SnX2MOs}).
The spectrum of \ce{SnI2}, on the contrary, does not exhibit this double-peak structure.
Peak C, still formed by the HOMO$\rightarrow$LUMO+1 transition, corresponds to an intense resonance at approximately 5.4 eV, while B is significantly weaker and downshifted (4.5 eV) compared to its counterpart in the spectra of the lighter monomers.
Between peaks B and C, an additional excitation with low OS appears: it comes from the transition between the HOMO-1 and the LUMO+2.
These variations compared to the spectra of \ce{SnCl2} and \ce{SnBr2} are attributed to the different energetic ordering and slightly modified spatial extension of the MOs contributing to the analyzed excitations (see Figures~\ref{fig:SnX2EnergyLevels} and \ref{fig:SnX2MOs}).

The effect of implicit solvation on the absorption spectra is consistent with the results obtained for the electronic structure.
In particular, peak A undergoes a blue shift in solution (see Figure~\ref{fig:SnX2AbsorptionSpectra}) due to the enhancement of the HOMO-LUMO gap already at the ground-state level (see Figure~\ref{fig:SnX2EnergyLevels}).
The same happens for peaks B and C, although the magnitude of the shift is much smaller in those cases.
It is worth noting that the solvatochromic shift of a certain excitation in the \ce{SnX2} molecule occurs regardless of the halide X.
The same trends are obtained also with the state-specific approach for the optical excitations~\cite{krumland-etal2021exploring}.

The absorption spectra calculated for the \ce{SnX2} molecules agree well with the limited experimental evidence available for these compounds. Absorption measurements of \ce{SnI2} solutions have revealed an onset of the first optical excitations in the range between 3 and 4~eV, depending on the solvent, which is in good agreement with theoretically calculated spectra~\cite{heo-etal2021enhancing,cao-etal2021stability}.
Furthermore, the characteristic shape of the spectrum with two distinct peaks matches experimental observations~\cite{cao-etal2021stability}.
Considering the corresponding lead dihalides, \ce{PbI2} solutions have been observed to have absorption maxima in a similar energy range below 4~eV~\cite{radicchi-etal2019understanding}.
Additionally, a methodologically similar computational study considering both implicitly and explicitly solvated \ce{PbX2} molecules reports UV-visible absorption spectra with a very similar energetic distribution of the individual transitions and the same trend of a red-shift of the absorption onset with larger halides~\cite{procida-etal2021first}.
Interestingly however, most of the calculated optical excitations of the \ce{PbX2} molecules are dark, with only a few bright excitations with large energetic separation, in strong contrast to our results for the \ce{SnX2} counterparts.

\section*{Summary and Conclusions}
In summary, we presented a first-principle study on the structural, electronic, and optical properties of tin halide monomers with chemical formula \ce{SnX_n^{2-n}}, with $n=1, 2, 3, 4$ and X = Cl, Br, I.
The bond lengths and angles of the relaxed geometries are found in agreement with reference values~\cite{hargittai2000molecular,levy-etal2003structure,neizer-etal2007vapor,kolonits-etal2004molecular}.
Partial charge analysis based on the Bader scheme reveals a significant electronic transfer from Sn to the halogen atoms, suggesting an ionic nature of the Sn-X bond with decreasing polarization in the presence of larger and less electronegative halogen species.
The charge displacement analysis supports these findings and additionally discloses more complex bond coordination, in agreement with earlier studies on metal-halide compounds~\cite{huber-etal2013directionality,inscoe-etal2021role,miertus-etal1981electrostatic,torii2019correlation,ramondo-etal1989molecular}.
The electronic structure of the neutral molecules is influenced by their composition and by the solvent polarization, both leading to variations of the HOMO-LUMO gap and, more generally, to the single-particle energy spectrum.
The real-space visualization of the MOs reveals that occupied orbitals have hybridized $s$- and $p$-character of both Sn and halogen atoms while the unoccupied states are primarily defined by $p$ atomic orbitals, consistent with previous findings on lead-halide counterparts~\cite{procida-etal2021first}.
The optical absorption spectra computed for the \ce{SnX2} monomers are all qualitatively similar, although the excitation energies vary significantly depending on the composition following the electronic-structure trends.
All of these molecules absorb in the UV region and are characterized by weak excitations at the onset, including the one due to the HOMO$\rightarrow$LUMO transitions.
Above 5~eV, stronger resonances disclosing distinct signatures in \ce{SnI2} compared to its lighter siblings.

In conclusion, our results provide insight into the fundamental quantum-mechanical characteristics of tin-halide molecules filling a gap in the existing literature and elevating the understanding of these compounds toward the level of lead halides, which have historically received much more attention.
As such, they offer a valuable reference for the growing community working on the synthesis, characterization, and prediction of lead-free halide perovskites.

\section*{Acknowledgements}
This work was supported by the German Research Foundation through the Priority Program SPP 2196 (Project number 424394788), by the German Federal Ministry of Education and Research (Professorinnenprogramm III), and by the State of Lower Saxony (Professorinnen f\"ur Niedersachsen and SMART). Computational resources were provided by the high-performance computing clusters CARL and ROSA at the University of Oldenburg, financed by the German Research Foundation (Projects No. INST 184/157-1 FUGG and INST 184/225-1 FUGG) and by the Ministry of Science and Culture of the State of Lower Saxony.
To avoid visual distortion or misrepresentation of data and increase accessibility for readers with color-vision deficiencies~\cite{crameri-etal2020misuse}, we made use of Scientific Colormaps designed by Crameri~\cite{crameri2023scientific}.


\section*{Data Availability}
The data that support the findings of this study are openly available in Zenodo with DOI 10.5281/zenodo.10374149.

\newpage

\providecommand{\latin}[1]{#1}
\makeatletter
\providecommand{\doi}
  {\begingroup\let\do\@makeother\dospecials
  \catcode`\{=1 \catcode`\}=2 \doi@aux}
\providecommand{\doi@aux}[1]{\endgroup\texttt{#1}}
\makeatother
\providecommand*\mcitethebibliography{\thebibliography}
\csname @ifundefined\endcsname{endmcitethebibliography}
  {\let\endmcitethebibliography\endthebibliography}{}


\end{document}